# Brief Announcement: Efficient Distributed Algorithms for Convolutional Neural Networks

Rui Li
lirui@cs.utah.edu
University of Utah
Salt Lake City, Utah, USA

Yufan Xu
yf.xu@utah.edu
University of Utah
Salt Lake City, Utah, USA

Aravind Sukumaran-Rajam
a.sukumaranrajam@wsu.edu
Washington State University
Pullman, Washington, USA

Atanas Rountev
rountev@cse.ohio-state.edu
Ohio State University
Columbus, Ohio, USA

P. Sadayappan
saday@cs.utah.edu
University of Utah
Salt Lake City, Utah, USA

## ABSTRACT

Several efficient distributed algorithms have been developed for matrix-matrix multiplication: the 3D algorithm, the 2D SUMMA algorithm, and the 2.5D algorithm. Each of these algorithms was independently conceived and they trade-off memory needed per node and the inter-node data communication volume. The convolutional neural network (CNN) computation may be viewed as a generalization of matrix-multiplication combined with neighborhood stencil computations. We develop communication-efficient distributed-memory algorithms for CNNs that are analogous to the 2D/2.5D/3D algorithms for matrix-matrix multiplication.

## CCS CONCEPTS

• **Theory of computation** → **Distributed algorithms**; **Parallel algorithms**.

## KEYWORDS

distributed algorithms; neural networks; communication optimization

## 1 INTRODUCTION

The design of efficient distributed-memory parallel algorithms is much more challenging than shared-memory parallel algorithms. A number of recent research efforts have focused on utilizing shared-memory parallelism for Convolutional Neural Networks (CNN) [3, 9, 11, 16, 17] and high-performance library implementations are also available from vendors, e.g., oneDNN from Intel [12] and cuDNN from Nvidia [4]. However, only very simple and restricted schemes have been implemented for distributed-memory parallel systems [1, 6, 10, 13]. This is in contrast to matrix-matrix multiplication, for which a number of communication-optimal distributed-memory parallel algorithms have been developed, including the 3D algorithm [2, 5, 7], the 2D SUMMA algorithm [15], and the 2.5D algorithm [14].

In this paper, we synthesize efficient distributed-memory algorithms for CNN:

$$Out[b,k,w,h] = In[b,c,\sigma_w w + r, \sigma_h h + s] * Ker[k,c,r,s]$$

where *Out* is the output feature map, *In* is the input feature map, *Ker* is the kernel; $b$ indexes the batch dimension, $k$ the output feature, $c$ the input feature, $h, w$ the vertical and horizontal pixel index, and $r, s$ the vertical and horizontal stencil index; $\sigma_w$ and $\sigma_h$ are strides.

Our key insight is that prior work on analytical modeling of data movement for tile-size optimization for CNN on shared-memory parallel systems [8] can be adapted for synthesis of communication efficient distributed-memory algorithms. The synthesis involves two steps: (1) formulation and solution of a 2-level tile-size optimization problem under a virtual global-memory model (Sec. 2.1), and (2) synthesis of initial data distribution and inter-processor collective communication schedule on a logical multi-dimensional processor view with a partitioned memory address space (Sec. 2.2).

```
for(b = 0; b < Nb; n++)
 for(h = 0; h < Nh; h++)
  for(w = 0; w < Nw; w++)
   for(k = 0; k < Nk; k++)
    for(c = 0; c < Nc; c++)
     for(r = 0; r < Nr; r++)
      for(s = 0; s < Ns; s++)
       Out[n][k][w][h] +=
        In[n][c][w+r][h+s]*Ker[k][c][r][s]
```
**Listing 1: CNN loops**

```
for(bt = 0; bt < Nb; bt+=Tb)
 for(ht = 0; ht < Nh; ht+=Th)
  for(wt = 0; wt < Nw; wt+=Tw)
   for(kt = 0; kt < Nk; kt+=Tk)
    for(ct = 0; ct < Nc; ct+=Tc)
     CNNTile(bt,ht,wt,kt,ct);
```
**Listing 2: CNN with single-level tiling**

## 2 BACKGROUND

Consider the CNN computation shown in Listing 1. Listing 2 shows a single-level tiled code for the computation, where five of the loops are tiled (excluding the stencil dimensions *Nr* and *Ns*, which are usually small prime numbers like 3, 5, 7 and are not tiled). In prior work [9], it was shown that the volume of data movement for the sequential execution of the tiled CNN code on a system with a single-level memory hierarchy could be analytically modeled as a function of the tile sizes and fast-memory capacity $M$ as:

$$\begin{aligned}
cost = &N_b N_k N_w N_h + N_k N_c N_r N_s N_w N_h N_b/(T_w T_h T_b)+ \\
&N_b N_c (\sigma_w T_w + N_r - 1)(\sigma_h T_h + N_s - 1)N_w N_h N_k/(T_w T_h T_k) \\
&(\sigma_w T_w + N_r - 1)(\sigma_h T_h + N_s - 1)T_b T_c + T_w T_h T_b T_k + \\
&N_r N_s T_k T_c \leq M, T_i \leq N_i, i \in \{b, k, w, h, c\}
\end{aligned} \quad (1)$$

The five tile loops in Listing 1 are fully permutable, and for each permutation, an analytical cost expression can be derived [9]. However, many permutations have exactly the same cost expression. Specifically, it can be shown (under some conditions) that the permutation of the outer four tile loops does not change the cost [9]. Further, the three indexes $b$, $h$, $w$ are equivalent with respect to the data reuse characteristics and can be treated as a composite index that ranges over the product of the three indices. In the treatment below, we will use $T_{bhw}$ to represent $T_b T_h T_w$, and these tile loops will always be treated as a contiguous band.

## 2.1 Parallel CNN: Global Virtual Memory

First, we consider data-movement optimization for CNN for a parallel system with $P$ processors, each with private local memory of capacity $M$, and a shared *virtual* global memory. Each processor can perform copy-in or copy-out of blocks of tensors to/from the virtual global memory to local memory. All computations are performed on data present in local memory, interspersed with data movement of blocks of data between local memory and the virtual global memory.

**Listing 3: Local loop schedule with c as the innermost tiling loop**

```
for  kt = 0:Wk/Tk
 for bt = 0:Wb/Tb
  for wt = 0:Ww/Tw
   for ht = 0:Wh/Th
    for ct = 0:Wc/Tc
     // load In, Ker tile from global memory;
     // when loading In, also load the "halo"
     for r = 0:Nr
      for s = 0:Ns
       for point loops k, b, w, h, c
        Out[bt*Tb+b, kt*Tk+k,
            wt*Tw+w, ht*Th+h] +=
        In[bt*Tb+b, ct*Tc+c,
           σw(wt*Tw+w)+r, σh(ht*Th+h)+s] *
        Ker[kt*Tk+k, ct*Tc+c, r, s]
     // store (update) Out tile to global memory
```

The CNN iteration space of size $N_b \times N_k \times N_c \times N_h \times N_w \times N_r \times N_s$ is partitioned into $P$ equal individual work-partitions of size $W_b \times W_k \times W_c \times W_h \times W_w \times N_r \times N_s$ per processor. The intersection of any pair of work-partitions is empty, and the union of all work-partitions covers the full iteration space. Each processor is responsible for executing the iteration-space points within its work-partition. But the available local memory may be insufficient to simultaneously hold all the data needed for executing its entire work-partition. Hence the work-partition is executed as a sequence of smaller $T_b \times T_k \times T_c \times T_h \times T_w \times N_r \times N_s$ tiles, whose data footprints fit within the available per-processor memory $M$. The work-partition and the iteration space are related as follows:

$$P * \prod W_i = \prod N_i, i \in \{b, c, k, h, w\} \quad (2)$$

We first address the following optimization problem: *Given a CNN with problem-size parameters $N_b$, $N_k$, $N_c$, $N_h$, $N_w$, $N_r$, $N_s$, and a parallel machine with $P$ processors with local memory $M$ per processor, find the optimal values for $W_b$, $W_k$, $W_c$, $W_h$, $W_w$, and $T_b$, $T_k$, $T_c$, $T_h$, $T_w$ so that the total volume of data moved between the virtual global memory and the local memories is minimized.*

Listing 3 shows one possible permutation for the tiled loops within a node; $a : b$ denotes the integers from $a$ to $b-1$ inclusive.

The total data movement volume for this loop is given by Eq. 3.

$$\begin{aligned}
cost = &W_b W_k W_w W_h + W_k W_c N_r N_s W_w W_h W_b/(T_w T_h T_b)+ \\
&W_b W_c (\sigma_w T_w + N_r - 1)(\sigma_h T_h + N_s - 1)W_w W_h W_k/(T_w T_h T_k) \\
s.t: &g = (\sigma_w T_w + N_r - 1)(\sigma_h T_h + N_s - 1)T_b T_c + T_w T_h T_b T_k + \\
&N_r N_s T_k T_c \leq M; 1 \leq T_i \leq W_i \leq N_i, i \in \{b, k, w, h, c\}; Equation\ 2
\end{aligned} \quad (3)$$

We proceed by formulating the following modified optimization problem that we can solve analytically, and use that solution to generate an efficient solution for Eq. 3. The main change from Eq. 3 to Eq. 4 is the simplification of the cost function and the memory capacity constraint by dropping the small $N_r - 1$ and $N_s - 1$ terms. Further, as discussed earlier, we replace $T_b T_h T_w$ by a new variable $T_{bhw}$ and $W_b W_h W_w$ by a new variable $W_{bhw}$. In addition, it is easy to see that an optimal solution would have $T_c = 1$, so we exclude this variable from Eq. 4. We denote the memory capacity in Eq. 4 by $M_L$; the relation between $M_L$ and the actual per-processor memory capacity will be established later.

$$\begin{aligned}
cost_L &= W_k W_{bhw} + \frac{N_k N_c N_{bhw}}{P}\left(\frac{N_r N_s}{T_{bhw}} + \frac{\sigma_w \sigma_h}{T_k}\right) \\
s.t: &g_L = T_{bhw} T_k \leq M_L; 1 \leq W_c \leq N_c; \\
&1 \leq T_i \leq W_i \leq N_i, i \in \{bhw, k\}; PW_{bhw} W_k W_c = N_{bhw} N_k N_c
\end{aligned} \quad (4)$$

We first observe that any optimal solution to Eq. 4 will satisfy

$$(W_k = T_k \text{ and } W_{bhw} = T_{bhw}) \text{ or } (W_c = N_c) \quad (5)$$

We must have either that $W_c = N_c$ or $W_c < N_c$. Suppose $W_c < N_c$ for some optimal solution. Consider the term $W_k W_{bhw}$ in the function to be optimized in Eq. 4. Since $W_c < N_c$ and $W_k W_c W_{bhw} P = N_k N_c N_{bhw}$, an increase in $W_c$ and a decrease in $W_k$ or $W_{bhw}$ could maintain this equality and could decrease the cost in Eq. 4, keeping all other variables the same. Thus, in an optimal solution with $W_c < N_c$ it must be the case that $T_k = W_k$ and $T_{bhw} = W_{bhw}$.

The following analysis considers two cases, based on Eq. 5.
**Case 1:** $W_c = N_c$
The arithmetic-mean geometric-mean inequality can be applied to find the following two solutions to the optimization problem in Eq. 4, depending on whether or not $M_L \leq N_k N_{bhw}/P$.
**Case 1a:** If $M_L \leq N_k N_{bhw}/P$, the optimal cost is achieved when $T_k = \sqrt{\frac{M_L \sigma_w \sigma_h}{N_r N_s}}$ and $T_{bhw} = \sqrt{\frac{M_L N_r N_s}{\sigma_w \sigma_h}}$. The cost is

$$cost^*_{L0} = N_k N_{bhw}/P + 2\frac{N_k N_c N_{bhw}}{P}\sqrt{\frac{N_r N_s \sigma_w \sigma_h}{M_L}} \quad (6)$$

| Condition | Cost |
|---|---|
| $N_k N_{bhw}/P \geq M_L$ | $N_k N_{bhw}/P + 2\frac{N_k N_c N_{bhw}}{P}\sqrt{\frac{N_r N_s \sigma_w \sigma_h}{M_L}}$ |
| $M_L \geq (\frac{N_k N_c N_{bhw}}{P})^{2/3}(N_r N_s \sigma_w \sigma_h)^{1/3}$ and $N_k N_{bhw}/P < M_L$ | $3(\frac{N_k N_c N_{bhw}}{P})^{2/3}(N_r N_s \sigma_w \sigma_h)^{1/3}$ |
| $M_L < (\frac{N_k N_c N_{bhw}}{P})^{2/3}(N_r N_s \sigma_w \sigma_h)^{1/3}$ and $N_k N_{bhw}/P < M_L$ | $M_L + 2\frac{N_k N_c N_{bhw}}{P\sqrt{M_L}}\sqrt{N_r N_s \sigma_w \sigma_h}$ |

Table 1: Summary of optimal solutions for Eq. 4 for tile loop permutations with $c$ as the innermost tiling loop.

| Condition | Cost |
|---|---|
| $N_k N_{bhw}/P \geq M_L$ and $N_r N_s N_k N_c/P \geq M_L$ and $\sigma_w \sigma_h N_c N_{bhw}/P \geq M_L$ | $\min(\frac{N_k N_{bhw}}{P}, \frac{N_k N_c}{P}, \frac{N_c N_{bhw}}{P})$ $+2\frac{N_k N_c N_{bhw}}{P}\sqrt{\frac{N_r N_s \sigma_w \sigma_h}{M_L}}$ |
| $M_L \geq (\frac{N_k N_c N_{bhw}}{P})^{2/3}(N_r N_s \sigma_w \sigma_h)^{1/3}$ and ($N_k N_{bhw}/P < M_L$ or $\sigma_w \sigma_h N_c N_{bhw}/P < M_L$ or $N_r N_s N_k N_c/P < M_L$) | $3(\frac{N_k N_c N_{bhw}}{P})^{2/3}(N_r N_s \sigma_w \sigma_h)^{1/3}$ |
| $M_L < (\frac{N_k N_c N_{bhw}}{P})^{2/3}(N_r N_s \sigma_w \sigma_h)^{1/3}$ and ($N_k N_{bhw}/P < M_L$ or $\sigma_w \sigma_h N_c N_{bhw}/P < M_L$ or $N_r N_s N_k N_c/P < M_L$) | $M_L + 2\frac{N_k N_c N_{bhw}}{P\sqrt{M_L}}\sqrt{N_r N_s \sigma_w \sigma_h}$ |

Table 2: Summary of optimal solutions for Eq. 4 considering all possible tile loop permutations.

**Case 1b:** If $M_L > N_k N_{bhw}/P$, the optimal cost is achieved when $T_k = \sqrt{\frac{N_k N_{bhw} \sigma_w \sigma_h}{P N_r N_s}}$ and $T_{bhw} = \sqrt{\frac{N_k N_{bhw} N_r N_s}{P \sigma_w \sigma_h}}$. The cost is

$$cost^*_{L1} = N_k N_{bhw}/P + 2\frac{N_k N_c N_{bhw}}{P}\sqrt{\frac{P N_r N_s \sigma_w \sigma_h}{N_k N_{bhw}}} \quad (7)$$

**Case2:** $T_k = W_k, T_{bhw} = W_{bhw}, W_c < N_c$

The Karush–Kuhn–Tucker conditions can be applied to find the following two solutions to the optimization problem in Eq. 4.

**Case2a:** If $M_L > N_k N_{bwh}$ and $M_L \geq (\frac{N_k N_c N_{bhw}}{P})^{2/3}(N_r N_s \sigma_w \sigma_h)^{1/3}$, the optimal cost is achieved when $T_k = (\frac{N_k N_c N_{bhw}}{P N_r N_s})^{1/3}(\sigma_w \sigma_h)^{2/3}$ and $T_{bhw} = (\frac{N_k N_c N_{bhw}}{P \sigma_w \sigma_h})^{1/3}(N_r N_s)^{2/3}$. The cost is

$$cost^*_{L2} = 3(\frac{N_k N_c N_{bhw}}{P})^{2/3}(N_r N_s \sigma_w \sigma_h)^{1/3} \quad (8)$$

**Case2b:** If $M_L > N_k N_{bwh}$ and $M_L < (\frac{N_k N_c N_{bhw}}{P})^{2/3}(N_r N_s \sigma_w \sigma_h)^{1/3}$, the optimal cost is achieved when $T_k = \sqrt{\frac{M_L \sigma_w \sigma_h}{N_r N_s}}$ and $T_{bhw} = \sqrt{\frac{M_L N_r N_s}{\sigma_w \sigma_h}}$. The cost is

$$cost^*_{L3} = M_L + 2\frac{N_k N_c N_{bhw}}{P\sqrt{M_L}}\sqrt{N_r N_s \sigma_w \sigma_h} \quad (9)$$

Combining the analysis from Case 1 and Case 2, and considering the relationship between different subcases, the solution to the optimization problem in Eq. 4 is summarized in Table 1.

The optimal solution to the modified optimization problem in equation (4) relates to the actual optimization problem of interest, Eq. 3, as follows:
- A valid efficient solution to Eq. 3 can be obtained by setting $M_L = M - \frac{1}{2}(3K(\sqrt{9K^2 + 4M} - 3K))$, where $K = \sqrt{\sigma_w \sigma_h N_r N_s}$.
- By setting $M_L = M$, the cost functions in Table 1 are lower bounds on the lowest possible costs for any valid tile sizes.

The analysis for other tile loop permutations is similar, and the results are presented in Table 2.

## 2.2 Distributed CNN: Partitioned Memory

In this subsection, we show how the tiled solution from Sec. 2.1 can be used to construct an efficient algorithm for CNN for a distributed-memory parallel system with a fully partitioned memory: each processor has a local memory capacity $M_D$ and can communicate with other processors only through explicit inter-processor communication operations.

With the global memory model used in Sec. 2.1, the local memory capacity in each processor ($M$) only needs to be sufficient to hold the data-footprints (slices of accessed data in the three tensors) of a tile, with blocks of data being moved between local memory and the virtual global memory in between successive tiles. In a distributed-memory parallel system, in addition to holding the data-footprints of the currently executed tile, all elements of the three tensors must also be held in the local memory of one or more processors.

We first provide a high-level sketch of the construction of the distributed-memory parallel CNN algorithm. Given CNN problem parameters $N_b, N_k, N_c, N_h, N_w, N_r, N_s$, number of processors ($P$), and memory $M_D$ per processor: i) Determine the per-memory capacity $M_T$ needed to hold the tensors in a distributed manner to compute available memory for tiles, $M = M_D - M_T$; ii) Use the reduced capacity $M$ to solve the global-memory optimization problem discussed in Sec. 2.1 – the same tile schedule will be used by the processors in the distributed-memory system; iii) Determine parameters $P_b, P_k, P_c, P_h, P_w$ to create a logical multi-dimensional grid for the $P$ processors; iv) Generate the initial data distribution and the data communication schedule for the multi-dimensional processor grid.

**Parameters for Multi-dimensional Processor Grid**: The solution to the global-memory optimization problem in Sec. 2.1 provides values for tile sizes $T_i$ as a function of problem parameters and machine parameters. For the distributed-memory CNN algorithms, the $P$ processors are viewed as a logical multidimensional $P_b \times P_h \times P_w \times P_c \times P_k$ grid, with $P_i = N_i/W_i$.

If the solution to the global-memory optimization problem addressed in Sec. 2.1 corresponds to **Case 2**, $P_k = N_k/W_k = N_k/T_k$, $P_b P_w P_h = N_b N_w N_h/(T_b T_w T_h)$, and $P_c = P/(P_k P_b P_w P_h)$.

For those solutions derived from **Case 1**, $P_c = 1, W_c = N_c, P_k = \frac{N_k}{W_k}, P_b P_w P_h = \frac{N_b N_w N_h}{W_b W_w W_h}, W_k = \sqrt{\frac{N_k N_b N_h N_w}{P}\frac{\sigma_w \sigma_h}{N_r N_s}}, W_b W_w W_h = \sqrt{\frac{N_k N_b N_h N_w}{P}\frac{N_r N_s}{\sigma_w \sigma_h}}$

The Case 1 solution is analogous to the 2D SUMMA [15] algorithm for distributed matrix-multiplication, and Case 2 corresponds to the 2.5D [14] and 3D [5] distributed matrix multiplication algorithms; the latter when $M_L \geq (\frac{N_k N_c N_{bhw}}{P})^{2/3}(N_r N_s \sigma_w \sigma_h)^{1/3}$ and the former otherwise.

**Initial Data Distribution:** For each tensor, one or more of the five loop indices $b, c, k, h, w$ are absent in the indexing expression: $k$ is absent in $In[b, c, \sigma_w w + r, \sigma_h h + s]$, $b, h, w$ do not appear in $Ker[k, c, r, s]$, and $c$ is not used in $Out[b, k, w, h]$. Therefore, identical data slices of a tensor will be accessed by all processors along any *missing* loop index. For example, identical slices of $Ker[k, c, r, s]$ will be accessed by processors that only differ in the processor-grid coordinates along $b$ or $h$ or $w$.

Each processor $P_{b,c,k,h,w}$ accesses a slice of $Ker$, where the specific elements accessed by a processor depend on its indices $c$ and $k$, but not on $b, h, w$. There are $P_c \times P_k$ distinct slices of data, each with $W_c \times W_k \times N_r \times N_s$ elements and accessed by $P_b \times P_h \times P_w$ different processors. The initial distribution of data for $Ker$ has each of these slices uniformly partitioned along $c$ into smaller sub-slices of size $\frac{W_c}{P_b P_h P_w} \times W_k \times N_r \times N_s$, distributed over the $P_b \times P_h \times P_w$ processors that need to access data in all the sub-slices. A similar initial distribution strategy is used for *In*, where the data slice needed by a processor is partitioned among $P_k$ processors. For *Out*, if $P_c > 1$, replication is used instead of disjoint partitioning across the $c$ dimension in the multi-dimensional processor grid, to avoid additional data movement compared to that required in the global-memory solution of Sec. 2.1.

**Collective Communication Schedule:** Each processor allocates buffers for storing the elements of *In* and *Ker* accessed by a tile. The sizes of buffers are $T_b(\sigma_w T_w + N_r - 1)(\sigma_h T_h + N_s - 1)$ for *In* tile and $T_k N_r N_s$ for *Ker* tile. There is no need for allocate any additional memory for *Out* tile because space for the entire accessed data slice for the tiles is allocated in the initial data distribution. There is therefore also no need for any inter-processor communication for *Out* (until a reduction step at the very end). Execution proceeds as a sequence of $W_c$ tiles with interspersed inter-processor communication between tiles.

The communication for *In* is as follows. The $W_c$ tiles are divided into $P_k$ groups, each group containing $W_c/P_k$ tiles. The first processor along the $k$ dimension of the processor grid broadcasts the elements of *In* needed for the tiles in the first iteration group. After $W_c/P_k$ steps, the next processor along the $k$ dimension becomes the originator of data broadcasts for the next $W_c/P_k$ steps, and so on. The communication schedule for *Ker* is analogous to *In*, with the broadcast of data being performed along the $b, w, h$ dimensions of the processor grid.

**Cost Analysis:** The communication cost of the above distributed CNN algorithm is as follows. Let $cost_I$ be the initialization cost (initialization cost for *In*, *Ker* and reduction cost for *Out*). It is equal to the footprint of the initial data distribution. The communication cost, $cost_C$, is equal to total volume of broadcast data for *In* and *Ker*. The total cost $cost_D = cost_C + cost_I$.

$$\begin{aligned}
cost_I =& W_b W_k W_w W_h + (\sigma_w N_w + N_r - 1)(\sigma_h N_h + N_s - 1)N_b N_c/P \\
& + N_r N_s N_k N_c / P \\
cost_C =& W_k W_c N_r N_s W_w W_h W_b/(T_w T_h T_b) + \\
& W_b W_c (\sigma_w T_w + N_r - 1)(\sigma_h T_h + N_s - 1) W_x W_y W_k/(T_w T_h T_k)
\end{aligned} \quad (10)$$

The local memory $M_D$ in each processor must hold the initial data layout for all three tensors, and the tile footprints for *Ker* and *In*, giving the following local memory constraint expression:

$$\begin{aligned}
g_D =& (\sigma_w T_w + N_r - 1)(\sigma_h T_h + N_s - 1)T_b T_c + N_r N_s T_k T_c \\
& + W_b W_k W_w W_h + N_r N_s N_k N_c/P \quad (11) \\
& + (\sigma_w N_w + N_r - 1)(\sigma_h N_h + N_s - 1)N_b N_c/P \le M_D
\end{aligned}$$

Let $cost, g$ be the *cost* and $g$ stated in Equation (3). We can see that both $cost_D - cost$ and $g_d - g$ equal $\frac{1}{P}(size(In) + size(Ker))$, which is a constant. Thus, the data movement cost of the distributed CNN algorithm only differs from that of the global-memory solution Table 2 by a constant.

## 3 CONCLUSION

This paper presents a new methodology to synthesize efficient distributed-memory algorithms. The key insight driving this work is that a two-level tile optimization model can be used to synthesize efficient distributed-memory algorithms. The methodology was used to create new distributed algorithms for convolutional neural networks, analogous to the well known 2D/2.5D/3D distributed matrix-multiplication algorithms.

## ACKNOWLEDGMENTS

This work was supported in part by the U.S. National Science Foundation through awards 1946752 and 1919122.